\documentclass[a4paper,twocolumn,pra,english,reprint, longbibliography, superscriptaddress, showkeys, showpacs=false, nofootinbib, hyphens]{revtex4}
\usepackage{color,footnote}
\usepackage{babel}
\usepackage{amsmath,amssymb}
\usepackage{mathtools,bbm}
\usepackage{subfigure}
\usepackage{graphicx}
\usepackage{physics,braket}
\usepackage{comment}
\usepackage{mathtools,mathrsfs}
\usepackage[varg]{txfonts}

\usepackage{ulem} 

\usepackage[unicode=true,pdfusetitle,bookmarks=true,bookmarksnumbered=false,bookmarksopen=false,breaklinks=true,pdfborder={0 0 0},backref=false,colorlinks=true]
 {hyperref}
\makeatletter
\@ifundefined{textcolor}{}
{%
 \definecolor{BLACK}{gray}{0}
 \definecolor{WHITE}{gray}{1}
 \definecolor{RED}{rgb}{1,0,0}
 \definecolor{GREEN}{rgb}{0,1,0}
 \definecolor{BLUE}{rgb}{0,0,1}
 \definecolor{CYAN}{cmyk}{1,0,0,0}
 \definecolor{MAGENTA}{cmyk}{0,1,0,0}
 \definecolor{YELLOW}{cmyk}{0,0,1,0}
}
\pdfoutput=1
\hypersetup{colorlinks=true,citecolor=blue,linkcolor=cyan,urlcolor=blue,filecolor= green, breaklinks=true}
\usepackage{url}

\newcommand{\mf}{\mathfrak}

\newcommand*{\img}[1]{%
    \raisebox{-.2\baselineskip}{%
        \includegraphics[
        height=\baselineskip,
        width=\baselineskip,
        keepaspectratio,
        ]{#1}%
    }%
}

\begin{document}

\author{J. S. Ara\'ujo\href{https://orcid.org/0000-0002-6086-7189}{\includegraphics[scale=0.05]{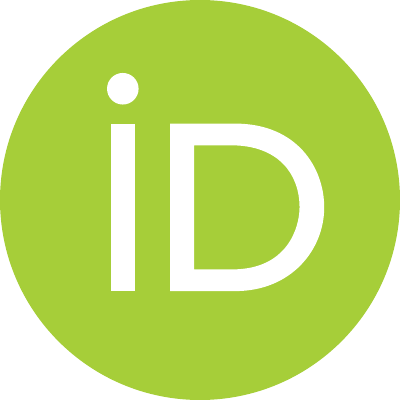}}}
\affiliation{Instituto de F\'isica, Universidade Federal do Rio de Janeiro,
Caixa Postal 68528, Rio de Janeiro, RJ 21941-972, Brazil}

\author{D. S. Starke\href{https://orcid.org/0000-0002-6074-4488}{\includegraphics[scale=0.05]{orcidid.pdf}}}
\affiliation{Physics Department, 
Federal University of Santa Maria, 97105-900,
Santa Maria, RS, Brazil}

\author{A. S. Coelho\href{https://orcid.org/0000-0002-1860-6743}{\includegraphics[scale=0.05]{orcidid.pdf}}}
\affiliation{Instituto de F\'isica, Universidade Federal do Rio de Janeiro,
Caixa Postal 68528, Rio de Janeiro, RJ 21941-972, Brazil}
\affiliation{Department of Mechanical Engineering, Federal University of Piauí, Teresina 64049-55, Piauí, Brazil}

\author{J. Maziero\href{https://orcid.org/0000-0002-2872-986X}{\includegraphics[scale=0.05]{orcidid.pdf}}}
\email{jonas.maziero@ufsm.br}
\affiliation{Physics Department, 
Federal University of Santa Maria, 97105-900,
Santa Maria, RS, Brazil}

\author{G. H. Aguilar\href{https://orcid.org/0000-0002-3457-506X}{\includegraphics[scale=0.05]{orcidid.pdf}}}
\email{gabo@if.ufrj.br}
\affiliation{Instituto de F\'isica, Universidade Federal do Rio de Janeiro,
Caixa Postal 68528, Rio de Janeiro, RJ 21941-972, Brazil}
\affiliation{Centro Brasileiro de Pesquisas Físicas, Rua Dr. Xavier Sigaud 150, 22290-180 Rio de Janeiro, Rio de Janeiro, Brazil}

\author{R. M. Angelo\href{https://orcid.org/0000-0002-7832-9821}{\includegraphics[scale=0.05]{orcidid.pdf}}}
\email{renato.angelo@ufpr.br}
\affiliation{Department of Physics, Federal University of Paran\'a, Curitiba, Paran\'a, P.O. Box 19044, 81531-980, Brazil}

\selectlanguage{english}

\title{Quantum observable's reality erasure with spacelike-separated operations}

\begin{abstract}
In 1935, Einstein, Podolsky, and Rosen argued that quantum mechanics is incomplete based on the assumption that local actions cannot influence elements of reality at a distant location (local realism). In this work, using a recently defined quantum reality quantifier, we show that Alice's local quantum operations can be correlated with the erasure of the reality of observables in Bob's causally disconnected laboratory. To this end, we implement a modified optical quantum eraser experiment, ensuring that Alice's and Bob's measurements remain causally disconnected. Using an entangled pair of photons and quantum state tomography, we experimentally verify that, even with the total absence of any form of classical communication, the choice of quantum operation applied by Alice on her photon is correlated with the erasure of a spatial element of reality of Bob's photon. Our results reveal that Bob's photon can entangle two extra non-interacting degrees of freedom, thus confirming that Bob's photon path is not an element of physical reality. 
\end{abstract}


\date{\today}

\maketitle

\section{Introduction}

Classical physics theories lie in harmony with the concepts of realism and local causality. Realism assumes that the physical properties of all systems are well defined regardless of any observer's interventions.
Local causality, on the other hand, suggests that a cause can only influence events that lie within its light cone. 
From these premises, it readily follows that physical reality can, by no means, be influenced at a distance (local realism). Quantum mechanics contradicts this framework. Such tension led Einstein, Podolsky, and Rosen (EPR) to challenge the theory’s completeness. Unwilling to negotiate the assumption of local causality, the authors put forward a rationale suggesting that noncommuting observables could be simultaneous elements of reality (EORs), which would imply quantum mechanics to be an incomplete model of the physical world~\cite{EPR1935}. 

%
%
Bohr’s complementarity countered this view, linking observable reality to experimental configurations ~\cite{Bohr1935}, thereby implying that noncommuting observables are incompatible. Modern experiments, known as quantum erasers~\cite{Scully1982,Kwiat1992,Herzog1995,Garisto1999,Walborn2002,Zubairy2004,Roos2004,Starke2023_2}, have shown that erasing welcher-weg information can restore wave-like behavior even when the choice of configuration is postponed in time~\cite{Jacques2007,Ma2016}. These results point to an active influence of the configuration choice on the system's behavior, thus corroborating Bohr's principle. In a different vein, Bell showed~\cite{Bell1964,Bell2001}, and numerous sophisticated experiments confirmed~\cite{Hensen2015,Giustina2015,Shalm2015,Hensen2016,Rauch2018,Li2018}, that nature is at odds with the hypothesis of local causality.
More recently, by employing Bohr's view of physical reality and reinforcing the perspective of nonlocality, a notable experiment was thus conducted which dismissed the premises of EPR's local realism simultaneously (i.e., reality and locality). In effect, Ma \textit{et al.}~\cite{Ma2013} demonstrated that the quantum erasure of the welcher-weg information at one location ``\textit{depends on the choice of measurement on the other (environment) photon, even when all the events on the two sides that can be space-like separated, are space-like separated.}'' In slightly different veins, several other intriguing works in the literature have explored related concepts, including the quantum eraser and various forms of quantum measurements, such as partial, non-local, or joint measurements \cite{Elitzur2001, Brodutch2016, Xu2019, Quintino2014,Uola2014}.

In this article, we deepen the EPR-Bohr debate by bringing into focus two novel elements. To present our proposal in connection with the aforementioned quantum erasure experiments, we start by asking the following: Can EORs at Bob's site be erased with causally disconnected operations made in Alice's laboratory?

The remainder of this article is organized as follows. In Sec.~\ref{sec:framework}, we present 
the quantum irreality quantifier we use to introduce the nonlocal quantum reality eraser.
Section~\ref{sec:experiment} describes our experimental setup, which implements a modified quantum eraser protocol using entangled photons and extra non-interacting degrees of freedom. In Sec.~\ref{sec:analysis}, we present the analytical predictions and derive the conditions under which quantum reality is erased at Bob's site. The experimental results are discussed in Sec.~\ref{sec:results}, followed by the conclusions in Sec.~\ref{sec:conclusion}.
Additional technical details and methodological clarifications are provided in the Appendixes~\ref{app:A} and~\ref{app:B}.

\section{Theoretical Framework}
\label{sec:framework}

%
%

Toward answering the question we raised in the Introduction, our first move consists of abandoning the wave-particle ontology. The motivation for this is provided by the quantum delayed-choice experiment reported in Ref.~\cite{Dieguez2022}, where it is shown that certain challenges to Bohr's complementarity principle can be overcome by use of the {\it irreality measure}~\cite{Bilobran2015}
\begin{equation}
	\mathfrak{I}_X(\rho)=S\big(\Phi_X(\rho)\big)-S(\rho) \label{eq:irr},
\end{equation}
where $S(\rho)=-\Tr(\rho \log_2 \rho)$ is the von Neumann entropy, $\rho$ is a quantum state acting on the Hilbert space $\mathcal{H=H_A\otimes H_B}$ (with $\mathcal{H_B}$ possibly multipartite), $\Phi_X(\rho)\coloneqq \sum_i\tilde{M}_i^x\rho\tilde{M}_i^x$ denotes a nonselective measurement of an observable $X=\sum_ix_iM_i^x$ acting on $\mathcal{H_A}$, and $\tilde{M}_i^x=M_i^x\otimes \mathbbm{1}_\mathcal{B}$ are projectors on the composite space. To introduce this measure, the authors of Ref.~\cite{Bilobran2015} considered a realism criterion according to which $X$ is said to be an EOR for a given state preparation $\rho$, when nonselective measurements of $X$ are innocuous, i.e., $\Phi_X(\rho)=\rho$. Only in this case, $\mathfrak{I}_X=0$ and $\rho$ is then called an $X$-reality state. When $\mathfrak{I}_X>0$, the measure indicates the extent to which the realism criterion is violated. The irreality of any observable $X$ is always non-negative, is upper-bounded by $\log_2(\dim{\mathcal{H_A}})$, and can be decomposed as $\mathfrak{I}_X(\rho)=C_X(\rho_\mathcal{A})+D_X(\rho)$.
Here $C_X(\rho_\mathcal{A})=S\big(\Phi_X(\rho_\mathcal{A})\big)-S(\rho_\mathcal{A})$ is the relative entropy of coherence~\cite{Baumgratz2014}, $D_X(\rho)$ is the quantum discord~\cite{Ollivier2001} of the $X$ measurement, and $\rho_{\mathcal{A}}=\Tr_{\mathcal{B}}(\rho)$. This shows that $X$ realism is prevented by two forms of quantumness, namely, quantum coherence and quantum correlations. Unlike classical realism, the framework of quantum (ir)realism prevents 
complementary observables $X$ and $X'$ from being EORs simultaneously, since $\mathfrak{I}_X(\rho)+\mathfrak{I}_{X'}(\rho)\geq I_\mathcal{A|B}(\rho)$, with the bound being the conditional information~\cite{Dieguez2022}.
Quantum irrealism has recently received much attention, both from the theoretical~\cite{Rudnicki2018,Dieguez2018,Freire2019,Lustosa2020,Savi2021,Gomes2022,Orthey2022,Basso2022,Paiva2023,Engelbert2023,Orthey2025,Orthey2025_2} and the experimental sides~\cite{Mancino2018,Dieguez2022}. Within a broader perspective, quantum irrealism also connects with further foundational elements \cite{Reid2017,Moreira2019,Zhu2021,Thenabadu2022,Starke2024}.

%
%

\begin{figure}[htbp]
\centering
\includegraphics[width=1\linewidth]{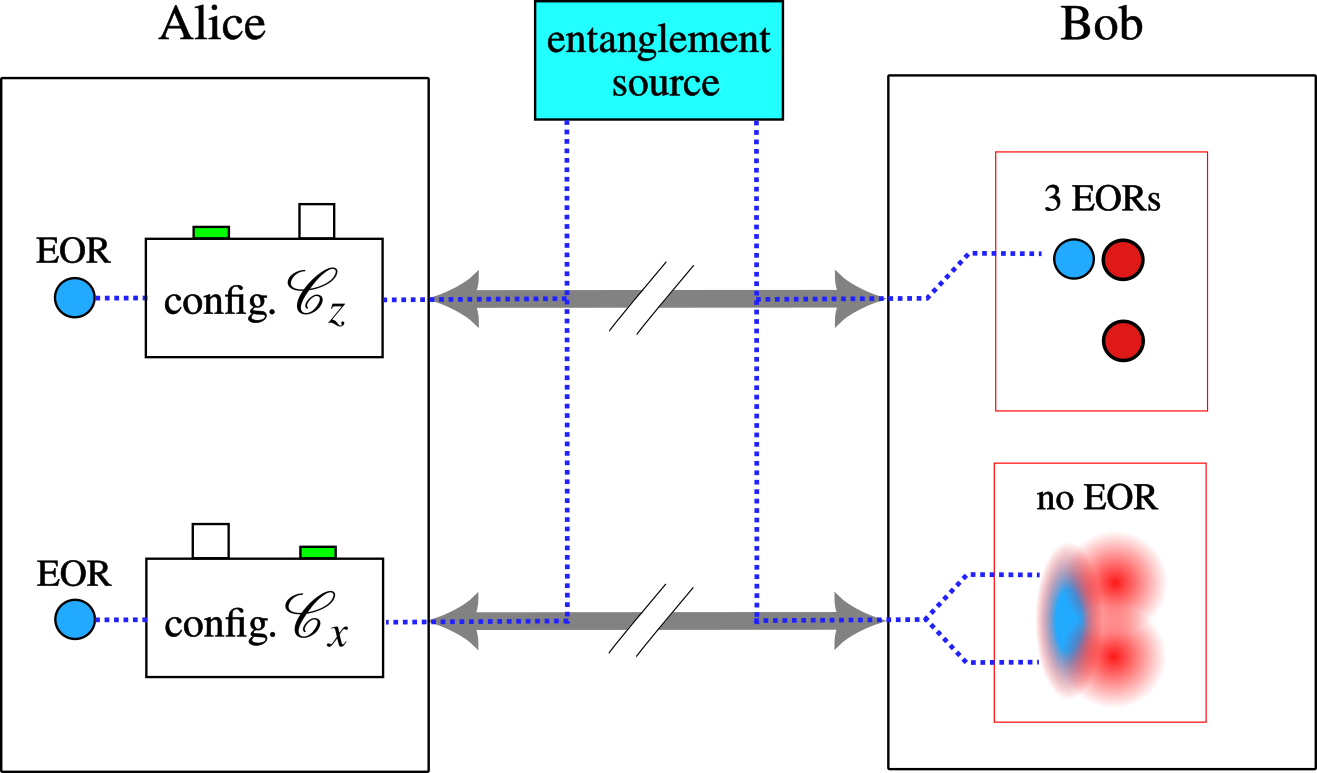}
\caption{Schematic of the {\it reality quantum eraser}. A pair of entangled photons is generated through beta-barium-borate crystals (correlation source), with one photon directed to Alice's site and the other to Bob's, where two d.o.f., $d_{1,2}$, have been prepared as EORs (solid red bullets), that is, $\mathfrak{I}_{d_{1,2}}=0$. Alice submits her photon to one of two configurations, $\mathscr{C}_{z(x)}$, and then measures the photon's path, assigning it an EOR (blue bullets on Alice's side). When Alice chooses the configuration $\mathscr{C}_z$ ($\mathscr{C}_x$) and measures her photon, Bob's photon acquires a spatial EOR (enters in path superposition), interacts with either $d_1$ or $d_2$ (interacts with both $d_{1,2}$), and leaves these d.o.f. correlated (entangled). Theoretical predictions show that, when $\mathscr{C}_x$ is chosen, the $d_{1,2}$ EORs are found erased after Alice's photon is measured, that is, $\mathfrak{I}_{d_{1,2}}>0$, even when the sites are spacelike separated.}
\label{fig1}
\end{figure}

As per the second distinctive element of our proposal, we introduce an altered version of the quantum eraser. Through this, the absence of spatial EORs (behavior hitherto referred to as wavelike) is diagnosed by the activation of entanglement between two non-interacting degrees of freedom (d.o.f), hereafter referred to as $d_{1,2}$. Although this adds complexity to the execution of the experiment, it allows us to demonstrate reality erasure without resorting to visibility measures, which are usually plagued with retro-inference~\cite{Angelo2015,Starke2023_1}. The schematic of our experiment is depicted in Fig.~\ref{fig1}. A pair of entangled photons is generated by an entanglement source
and each photon of the pair is sent to a specific location. Meanwhile, in Bob's site, $d_{1,2}$ are prepared as EORs (red bullets). Alice submits the received photon to one of two different configurations, $\mathscr{C}_{z(x)}$ (notation to be clarified later), and measures her photon's path, which then becomes an EOR. Polarization is also projected, but it is not represented. When Alice chooses $\mathscr{C}_z$ and makes the measurements, Bob's photon path manifests as an EOR. Next, the photon interacts with either $d_1$ or $d_2$, which have their EORs preserved. However, when Alice chooses $\mathscr{C}_x$, Bob's photon path becomes indefinite (no longer EOR), the photon interacts with both $d_1$ and $d_2$, and the three d.o.f. end up entangled. As previously mentioned, quantum correlations activate the irrealism of $d_{1,2}$. The main result of this work is to experimentally demonstrate that this reality erasure occurs even when Alice's operations are spacelike-separated from Bob's site.

%
%

\begin{figure*}[htb]
\centering
\includegraphics[scale=0.28]{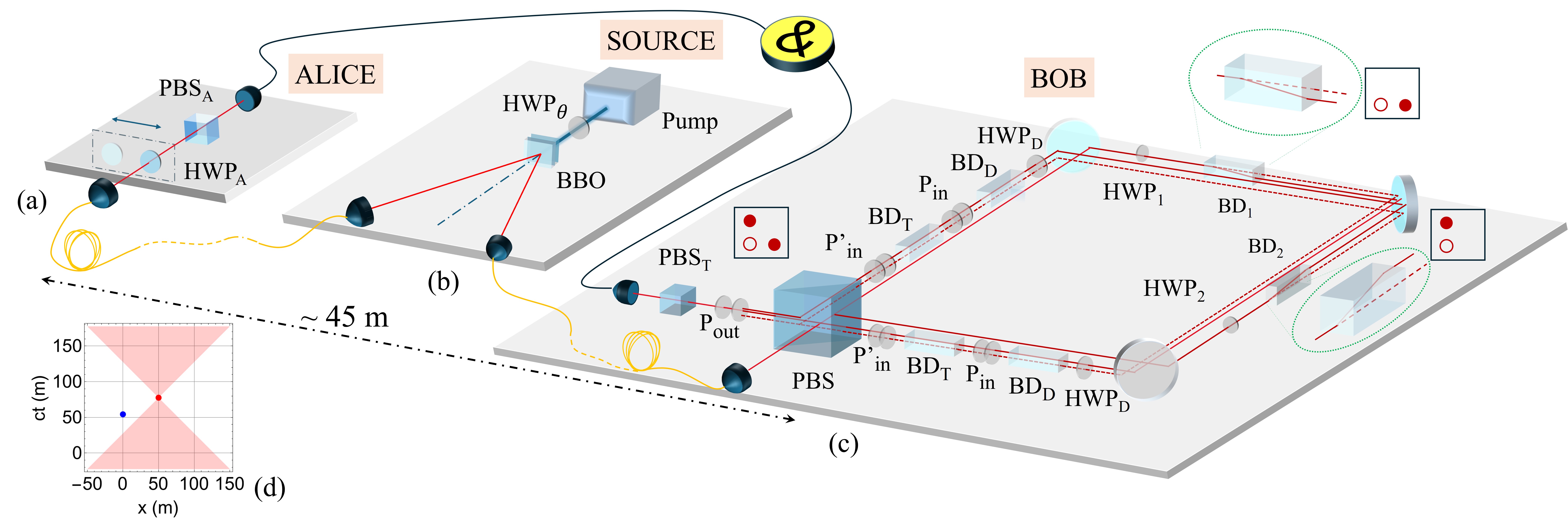}
\caption{A photon pair source located in the middle produced polarization-entangled photons at center wavelengths of $650\;\text{nm}$. These photons were then routed through optical fibers and transmitted to separate laboratories belonging to Bob and Alice, situated $45\;\text{m}$ apart from each other. The inset (d) shows the space-time location of the events corresponding to Alice's (red point) and Bob's (blue point) measurements in the performed experiments. Thus, Alice's quantum operations are space-like separated from Bob's measurements. Alice is free to decide whether to insert HWP$_\text{A}$ into the photon’s path before directing it to PBS$_\text{A}$. In Bob's laboratory there is a Sagnac interferometer with the clockwise ($\ket{0}_b$) and counterclockwise ($\ket{1}_b$) path accessed, respectively, by photons transmitted and reflected in the PBS. The beam-displacer BD$_1$ (BD$_2$) introduces an extra d.o.f. $d_1$ ($d_2$) in the $\ket{1}_b$ ($\ket{0}_b$) path. The sets P$_{\text{in}}$, P'$_{\text{in}}$ and BD$_\text{T}$ constitute the apparatuses used to perform the quantum state tomography related to $d_{1,2}$, while the sets PBS$_\text{T}$ and P$_{\text{out}}$ perform the quantum state tomography related to $b$. The measurements are performed utilizing avalanche photodetectors with a coincidence acquisition device.}
\label{fig:expsetup}
\end{figure*}

\section{Experimental setup}
\label{sec:experiment}

The experimental setup is illustrated in Fig.~\ref{fig:expsetup}.
Using type-I spontaneous parametric down-conversion in BBO crystals, we generate entangled photon pairs ~\cite{kwiat99} [Fig.~\ref{fig:expsetup}~(b)]. For more technical information about the implementation of the experimental setup, consult the Appendix~\ref{app:A}. The produced state is of the form $\ket{\Psi_+}_{AB}=\mf{c}\ket{00}_{AB} + \mf{s}\ket{11}_{AB}$, where $\ket{0}_A$ and $\ket{1}_A$ ($\ket{0}_B$ and $\ket{1}_B$) represent the horizontal and vertical polarizations of the photons sent to Alice (Bob). The probability amplitudes $\mf{c}$ and $\mf{s}$ are controlled by the half-wave plate HWP$_\theta$ angle, where $\mf{c}=\cos(\frac{\theta}{2})$, $\mf{s}=\sin(\frac{\theta}{2})$, and $\theta$ is the angle between the fast-axis of the plate and the horizontal polarization. Subsequently, the photons are sent through single-mode optical fibers to the Alice and Bob laboratories, which are spatially separated by a distance of more than $45$ meters, ensuring that, for each photon pair, the quantum operations executed on Alice's photon and Bob's photon are causally disconnected [see Fig.~\ref{fig:expsetup}~(d)]. At her site, Alice can decide whether to adjust a half-wave plate (HWP$_\text{A}$) to $0^\circ$ or $22.5^\circ$ before sending the photon to a polarizing beam splitter, PBS$_\text{A}$, that completely transmits (reflects) light with horizontal (vertical) polarization [Fig.~\ref{fig:expsetup}~(a)]. The two angles of HWP$_\text{A}$ represent a choice of the projection onto which Alice's measurement occurs: at $0^\circ$, the polarization is projected onto $\ket{0}_A$ (an eigenstate of $\sigma_z$), whereas at $22.5^\circ$, the polarization is projected onto $\ket{+}_A=\frac{1}{\sqrt{2}}(\ket{0}_A+\ket{1}_A)$ (an eigenstate of $\sigma_x$). Accordingly, the configurations will be referred to as $\mathscr{C}_z$ or $\mathscr{C}_x$, respectively. The photon sent to Bob's laboratory is directed into a modified Sagnac interferometer [Fig.~\ref{fig:expsetup}~(c)]. Upon incidence on the PBS, the component of light with horizontal (vertical) polarization travels through the interferometer clockwise (counterclockwise). These d.o.f. are identified as $\ket{0}_b$ for the clockwise path and $\ket{1}_b$ for the counterclockwise path. In each path $b$ of Bob's interferometer, three beam displacers (BD) are placed, which deviate the light depending on the incident polarization and the orientation of its optical axis: BD$_1$ (BD$_2$) deviates the horizontal (vertical) polarization horizontally, \img{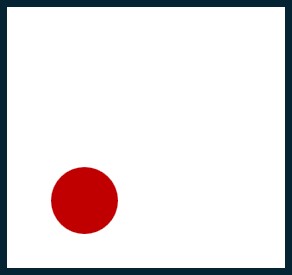}~$\longrightarrow$~\img{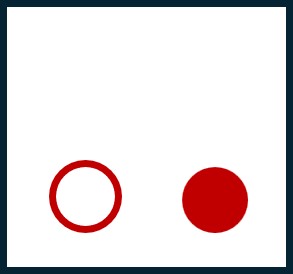} \big(vertically,  \img{1mode.jpg}~$\longrightarrow$~\img{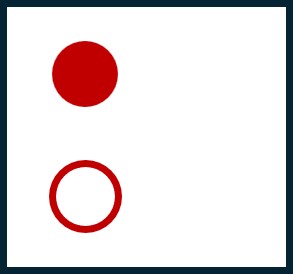}\big) [see the inset of Fig.~\ref{fig:expsetup}~(c)]. Thus, BD$_{1,2}$ implement operations on two extra d.o.f., with corresponding states $\ket{i}_{d_{1,2}}$ ($i=0,1$), controlled by the polarization of incident light. The notation is such that $\ket{1}_{d_{1,2}}$ encodes the undeviated initial paths. Upon traversing the beam-displacers BD$_{1,2}$, the photon's path gets entangled with the polarization, depending on the orientation of HWP$_{{1,2}}$, respectively. In general, the resulting state after BD$_k$ is $\delta_k \ket{0}_{B} \ket{0}_{d_k}+\gamma_k\ket{1}_{B} \ket{1}_{d_k}$ ($k=1,2$), which means a transformation from \img{1mode.jpg} to \img{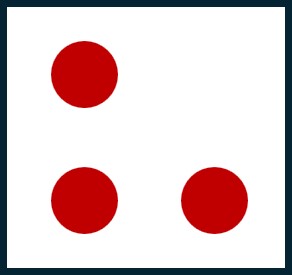} on the output of the Sagnac interferometer. Initially, the state of the composite system is given by $\ket{\Psi_0}=\ket{\Psi_+}_{AB}\ket{00}_{ab}\ket{11}_{d_1d_2}$, meaning that the photons' paths $a$ and $b$, and the extra d.o.f. $d_{1,2}$ are all EORs, while the polarizations $A$ and $B$ are not. With these principles in mind, we proceed to describe the state in two emblematic stages of the experiment.

\section{Theoretical Predictions}
\label{sec:analysis}
 
After Bob's photon enters the Sagnac interferometer and passes through the PBS, the state of the system results in
\begin{equation}
\ket{\Psi_1}=\big(\mf{c}\ket{000}_{ABb}+\mf{s}\ket{111}_{ABb}\big)\ket{0}_a\ket{11}_{d_1d_2}.
\label{eq:psi1}
\end{equation}
The phases resulting from reflections are not considered because  they can be compensated by tuning the angle of the beam displacers. From now on, we will assume that the plates HWP$_{1}$ and HWP$_{_2}$ are set to $\gamma_1=\delta_2=0$. After interacting with the beam displacers BD$_{1,2}$, the state transforms to
\begin{equation}
\ket{\Psi_2}=\big(\mf{c}\ket{000}_{ABb}\ket{01}_{d_1d_2}+\mf{s}\ket{111}_{ABb}\ket{10}_{d_1d_2}\big)\ket{0}_a.
\label{eq:psi2}
\end{equation}
Equations~\eqref{eq:psi1} and \eqref{eq:psi2} are subsidiary to the analysis that follows. We are now ready to assess how Alice's choice of configuration $\mathscr{C}_z$ or $\mathscr{C}_x$ correlates with the realism of the d.o.f. $\{b,d_1,d_2\}$ accessible to Bob. To compute the irreality of each of these d.o.f., we perform quantum state tomography after Alice and Bob perform projections through the following scheme: $\mathscr{C}_z$ and PBS$_\text{A}$ project the d.o.f. of Alice's photon onto $\ket{00}_{Aa}$, whereas $\mathscr{C}_x$ and PBS$_\text{A}$ project to $\ket{+0}_{Aa}=\frac{1}{\sqrt{2}}(\ket{0}_A+\ket{1}_A)\ket{0}_a$. Bob, in turn, always performs a projection onto the state $\ket{+}_B= \frac{1}{\sqrt{2}}(\ket{0}_B+\ket{1}_B)$.

Let us first consider the $\mathscr{C}_z$ configuration. The projections transform the state $\ket{\Psi_1}$ in Eq.~\eqref{eq:psi1} to $\ket{000011}_{ABabd_1 d_2}$, so that the state accessible in Bob's location is $\Omega_1^z\equiv\ket{\Omega_1^z}\bra{\Omega_1^z}$, with $\ket{\Omega_1^z}=\ket{011}_{bd_1 d_2}$.
The application of the dephasing map $\Phi_b$ is innocuous in this situation, i.e., $\Phi_b(\Omega_1^z)=\Omega_1^z$, resulting in $\mf{I}_b(\Omega_1^z)=0$.
Therefore, by post-selecting on $\ket{0}_a$, Alice guarantees that path $b$ is an EOR, meaning that Bob will not be able to verify quantum irreality for the photon sent to his laboratory. In the $\mathscr{C}_x$ configuration, the projections transform the state $\ket{\Psi_1}$ into
$\ket{+\!+\!0\beta_+11}_{ABabd_1 d_2}$, where $\ket{\beta_+}_{b}\equiv\mf{c}\ket{0}_{b}+\mf{s}\ket{1}_{b}$.
Hence, the state accessible to Bob is $\Omega_1^x=\ket{\Omega_1^x}\bra{\Omega_1^x}$. Now comes the crux: because the dephasing map $\Phi_b$ destroys the path coherence in $\ket{\beta_+}_b$, one has $\Phi_b(\Omega_1^x)\ne\Omega_1^x$ and $\mf{I}_b(\Omega_1^x)~=~-\mf{c}^{2}\log_{2}\mf{c}^{2}-\mf{s}^{2}\log_{2}\mf{s}^{2}$.
The irreality depends on the parameters that determine the initial entanglement between $A$ and $B$. In particular, maximum entanglement $\big(\mf{c} = \mf{s} =\frac{1}{\sqrt{2}}\big)$ triggers maximum irreality, which means that there is no EOR for Bob's photon path. 

\begin{figure}[htbp]
	\centering
	\includegraphics[scale=0.63]{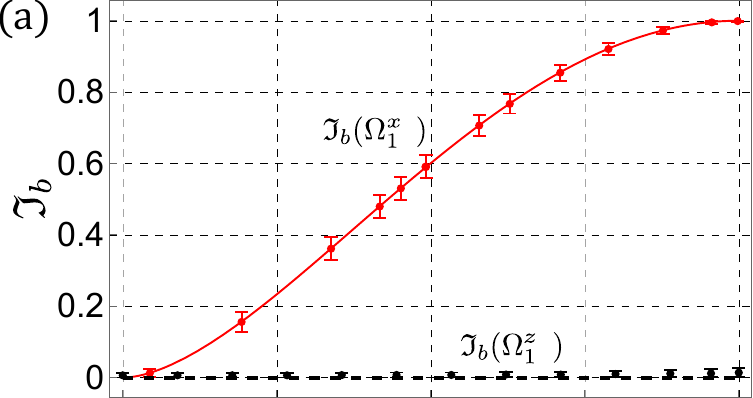}\\
	\vspace{0.2cm}
	\includegraphics[scale=0.63]{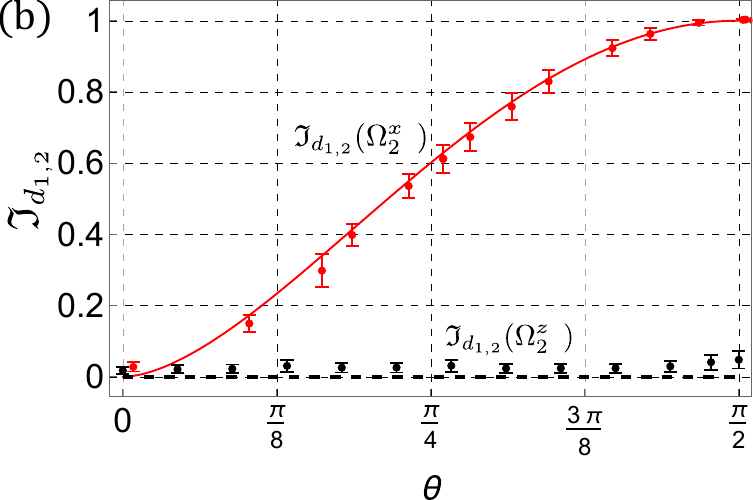}
	\caption{Theoretical (solid and dashed lines) and experimental (points) results for the irreality as a function of $\theta$, the parameter related to the initial polarization entanglement; the higher $\theta$ the stronger the entanglement. (a) Irreality $\mf{I}_b$ of the photon path $b$ calculated from quantum state tomography applied before Bob's photon interacts with BD$_{1,2}$. (b) Irrealities $\mf{I}_{d_{1,2}}$ the d.o.f. $d_{1,2}$, with $\mathfrak{I}_{d_1}=\mathfrak{I}_{d_2}$, calculated for the quantum state obtained after the interaction with the BD$_{1,2}$. In plots (a) and (b), the dashed line and the corresponding data points represent the scenario in which Alice selects the $\mathscr{C}_z$ configuration. Conversely, the solid line and the corresponding data points represent the $\mathscr{C}_x$ configuration.    }
	\label{fig:irrealismo1}
\end{figure}

So far, we have shown that the EORs of $b$ are erased in a one-to-one connection with Alice's choice. We now show that $b$'s irreality can be witnessed without resorting to visibility measurements and retro-inference. Further insights and a straightforward MZI application of our method for retro-inference prevention are presented in Appendix~\ref{app:B}. The rationale is that when $b$ is an EOR, thus having spatial localization, the d.o.f. $d_{1,2}$ cannot enter into superposition. Let us first analyze the configuration $\mathscr{C}_z$. After the projection operations, the state $\ket{\Psi_2}$ transforms into $\ket{000001}_{ABabd_1 d_2}$. In this case, the state accessible to Bob is $\Omega_2^z=\ket{\Omega_2^z}\bra{\Omega_2^z}$, where $\ket{\Omega_2^z}=\ket{001}_{bd_1d_2}$. The action of maps $\Phi_{d_{1,2}}$ is innocuous because qubits $d_{1,2}$ are neither entangled nor in superposition. It follows that $\mf{I}_{d_{1,2}}(\Omega_2^z)=0$, which means that $d_{1,2}$ are EOR.
Consider now the configuration $\mathscr{C}_x$. Performing the pertinent projections over $\ket{\Psi_2}$, Bob is granted the state $\Omega_2^x=\ket{\Omega_2^x}\bra{\Omega_2^x}$, where $\ket{ \Omega_2^x}=\mf{c}\ket{001}_{bd_1d_2}\pm\mf{s}\ket{110}_{bd_1d_2}$. As a consequence, $\mf{I}_{d_{1,2}}(\Omega_2^x)=-\mf{c}^{2}\log_{2}\mf{c}^{2}-\mf{s}^{2}\log_{2} \mf{s}^{2}$, implying that $d_{1,2}$ will no longer be EORs in general.
In particular, the irreality values are numerically equal to the amount of entanglement encoded in both the initial state $\ket{\Psi_0}$ and the state $\ket{\Psi_2^x}$.  

We have thus far shown that the configuration chosen by Alice correlates with the $b$ realism in Bob's site and that Bob can verify this effect by examining the quantum irreality (supported by entanglement) of the d.o.f. $d_{1,2}$, regardless of the distance between Alice's and Bob's sites. For space-like separated sites, one cannot, of course, claim that Alice controls reality at a distance through any mechanism consistent with relativistic causality. In fact, there exists a reference frame in which Alice's operations occur after Bob's.
Still, quantum reality erasure can be observed in full correlation with causally disconnected operations, as we now experimentally demonstrate.

\section{Experimental Results}
\label{sec:results}

We now report our experimental results for each of the four scenarios mentioned earlier. Our procedure consisted of projecting the d.o.f. $\{A,a,B\}$ and performing quantum state tomography procedures for the states encoding the d.o.f. $\{b,d_1,d_2\}$. The HWP$_\text{D}$ and BD$_\text{D}$
in Fig.~\ref{fig:expsetup}~(c) project Bob's photon polarization onto the basis $\{\ket{0}_B,\ket{1}_B\}$. To perform the quantum state tomography protocol regarding $d_{1,2}$, pairs of plates P$_{\text{in}}$ and P'$_{\text{in}}$ are used, composed of a HWP and a quarter-wave plate (QWP). These apparatuses, combined with the beam displacers BD$_\text{T}$ and the PBS of the interferometer, form the devices for quantum state tomography. The pair P${_\text{out}}$ and PBS$_\text{T}$ inserted in the output of the interferometer is responsible for the tomography related to the path $b$.
The adopted quantum state tomography protocol followed the description of Ref.~\cite{James2001}.

Let us first consider the state $\ket{\Psi_1}$ in Eq.~(\ref{eq:psi1}), a scenario in which Bob's photon does not interact with the beam displacers BD$_{1,2}$ and freely traverses the interferometer. In Fig.~\ref{fig:irrealismo1}~(a), the irreality of the photon path $b$ is presented, calculated from density matrices measured by quantum state tomography realized for both configurations $\mathscr{C}_{z(x)}$ in Alice's laboratory. Similarly, Fig.~\ref{fig:irrealismo1}~(b) shows the irreality of  $d_{1,2}$ (d.o.f. that underwent interaction with the beam displacers BD$_{1,2}$) calculated by tomography of the state $\ket{\Psi_2}$ [Eq.~\eqref{eq:psi2}]. 
For all the data, the error bars were calculated using a Monte Carlo simulation and indicate the uncertainty range within the experimental data. In both scenarios, the parameter $\theta$ governs the extent of entanglement within the state produced by the source. When $\theta=0$, the initial state $\ket{\Psi_+}_{AB}=\ket{00}_{AB}$ is entirely separable, yielding an irreality value equal to zero. As $\theta$ increases, irrealism monotonically increases, culminating at its peak when $\theta=\pi/2$. This corresponds to the scenario where one initially has the maximally entangled state $\ket{\Psi_+}_{AB}=\frac{1}{\sqrt{2}}(\ket{00}_{AB}+\ket{11}_{AB})$. It is then natural to conclude that the entanglement produced by the photon pair source is the fundamental resource behind the phenomenon being investigated. In reference to the decomposition $\mathfrak{I}_X(\rho)=C_X(\rho_\mathcal{A})+D_X(\rho)$, it is noteworthy that the irrealism has two responsible distinct mechanisms for Figs.~\ref{fig:irrealismo1}~(a)~and~(b) when Alice chooses $\mathscr{C}_x$. In Fig.~\ref{fig:irrealismo1}~(a), the irrealism associated with the path $b$ arises from quantum coherence. In this instance, path $b$ represents a delocalized system, which is the sole responsible for the tripartite entanglement generated for the triple $\{b,d_1,d_2\}$. In turn, it is precisely this inseparability (manifested as measurement discord $D_{d_{1,2}}$) that prevents $d_{1,2}$ from being EORs.

\section{Conclusions}
\label{sec:conclusion}

In conclusion, we proposed and implemented a quantum optical experiment involving six qubits to demonstrate nonlocal correlations between Alice's operations and quantum reality erasure at Bob's remote site. Our modified quantum eraser experiment advances previous setups in two fundamental ways. First, we abandon the wave-particle dichotomy in favor of quantum irrealism. This has the advantage of considering Bohr's complementarity in an updated way~\cite{Dieguez2022}, in opposition to EPR's notion of EORs. Second, by introducing extra d.o.f. (two noninteracting qubits), we manage to demonstrate the erasure of spatial EORs without resorting to typical retro-inference arguments. Our results thus provide strong evidence against local realism. 
We leave open the question of which quantum resource in the initially shared state is necessary for the nonlocal quantum reality erasure to occur.

\begin{acknowledgments}
We acknowledge the support of the National Institute for the Science and Technology of Quantum Information (INCT-IQ), Grant No. 465469/2014-0, the National Council for Scientific and Technological Development (CNPq), Grants No. 305957/2023-6, No. 309862/2021-3,No. 409673/2022-6, and PQ No. 305578/2023-5, the Coordination for the Improvement of Higher Education Personnel (CAPES), Grant No. 88887.827989/2023-00, and the John Templeton Foundation Grant No. 62424. GHA acknowledges the Funda\c{c}\~ao Carlos Chagas Filho de Amparo \`{a} Pesquisa do Estado do Rio de Janeiro (FAPERJ) JCNE Grant No. E-26/201.355/2021, and the  Funda\c{c}\~ao de Amparo \`{a} Pesquisa do Estado de S\~{a}o Paulo (FAPESP) Grant No. 2021/96774-4. We thank Luiz C. Ryff, Alexandre D. Ribeiro, Felipe E. L. da Cruz, Douglas F. Pinto, Gabriela B. Lemos, and Jacques Pienaar for helpful discussions.
\end{acknowledgments}

\appendix

\section{Technical details of the experimental implementation}
\label{app:A}

Figure \ref{fig:expsetup} depicts our experimental setup. A 50 mW continuous-wave He-Cd laser, centered at 325 nm and FWMD equal to 5 nm, pumps two BBO crystals in the cross-axis configuration. Through the process of spontaneous parametric down-conversion, pairs of photons are created, centered at 650 nm, in a state $\ket{\Psi_+}_{AB}=\mf{c}\ket{00}_{AB} + \mf{s}\ket{11}_{AB}$. The probability amplitudes $\mf{c}$ and $\mf{s}$ satisfy the normalization condition given by $|\mf{c}|^2+|\mf{s}|^2=1$ and can be manipulated using HWP$_{\theta}$. Due to the low optical power of the pump, the generation of multiple pairs of photons, such as the creation of four or six photons per coherence time of the laser, is neglected. After coupling the photons into single-mode fibers, they are sent to Alice and Bob's laboratories, which are separated by 45 m. The optical fibers implement a unitary operation, and a couple of HWPs and QWPs (not shown) compensate for this by applying the inverse of that operation, correcting any disturbances in the polarization of the single photons. After this correction, the purity of the biphoton state is preserved at 0.96, and the equations in the main text accurately describe the quantum states of the photons. In Alice and Bob's photon detection schemes, we use two silicon avalanche photodiodes ($\tau$-SPAD-FAST, PicoQuant) with an efficiency of $0.60 \pm 0.05 \%$ for 650 nm wavelength photons, dark counts below 400 per second, and a dead time of 60 ns. 

In her laboratory, Alice can perform projections onto the state $\ket{0}_A$ ($\ket{+}_A$ and $\ket{-}_A$) when adjusting the HWP$_A$  to $0^\circ$ ($22.5^\circ$). In our sketch, we illustrate this choice by depicting a round glass placed in and out of the photons' path. In Bob's station, the photons pass through a modified Sagnac interferometer that couples the polarization degrees of freedom with the path, as explained in the main text. The photons arrive at a PBS that transmits horizontal polarization while reflecting vertical polarization. Thus, photons originally in the polarization states $\ket{0}_B$ and $\ket{1}_B$ travel through the interferometer clockwise and anticlockwise, represented by the states $\ket{0}_b$ and $\ket{1}_b$, respectively. The photons traverse three beam displacers in each path, which transmit or deviate the photons depending on their polarization, as explained in the main text. Depending on the orientation angle of HWP$_{1(2)}$, we obtain an entangled state between the polarization and the path given by $\delta_k\ket{0}_B\ket{0}_{d_k}+\gamma_k\ket{1}_B\ket{1}_{d_k}$, with $k=1(2)$ representing the clockwise (anticlockwise) path. The optical elements HWP$_\text{D}$, BD$_\text{D}$, P$_{\text{in}}$, BD$_\text{T}$, P$^{\prime}_{\text{in}}$, PBS, and P$_{\text{out}}$, perform quantum state tomographic projections onto the relevant d.o.f. of the photons.

\begin{figure*}[h]
\centering
\includegraphics[scale=0.63]{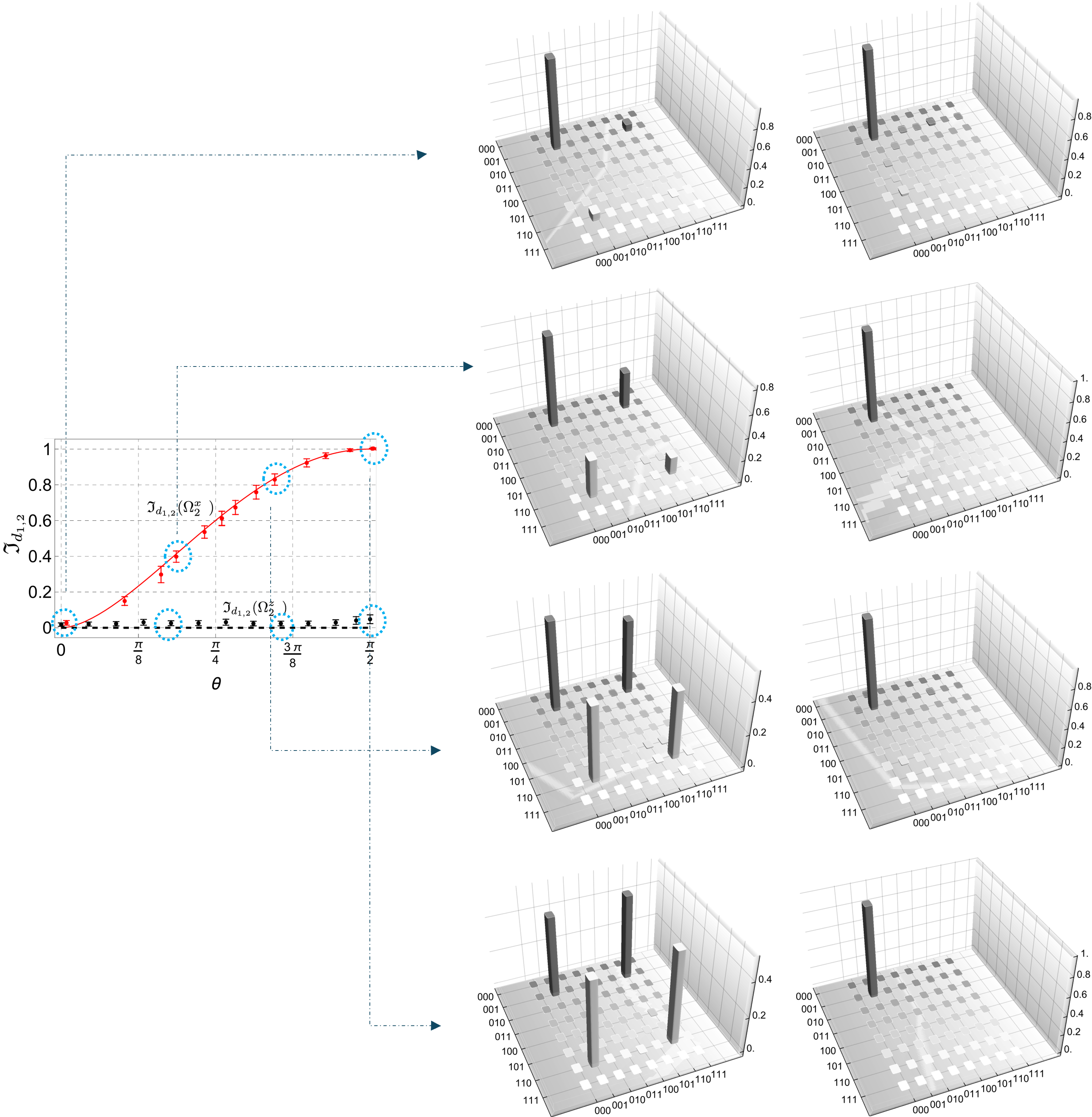}
\caption{Density matrices of the Hilbert space $\{b,d_1,d_2\}$ obtained for different values of $\theta$. The blue dashed circles in the graph of $\mf{I}_{d_{1,2}}$ indicate the corresponding points for each density matrix. The left (right) column corresponds to the matrices obtained when Alice implements projections onto $\ket{+}$ ($\ket{0}$).}
\label{fig:density}
\end{figure*}

In Fig.~\ref{fig:density}, we present the results of the real part of the reconstructed density matrices obtained for different values of $\theta$. Due to the negligible values of the imaginary part compared to the real part, we show only the results for the latter. This absence of significant imaginary components arises from the specific characteristics of our experimental state preparation and measurement protocol: (i) The entangled photon pairs are generated via spontaneous parametric down-conversion (SPDC) in a cross-axis configuration using two BBO crystals. The resulting state $\ket{\Psi_+}_{AB}=\mf{c}\ket{00}_{AB} + \mf{s}\ket{11}_{AB}$ is engineered to have no relative complex phase between the \(\ket{00}\) and \(\ket{11}\) components; (ii) all optical elements in the setup implement near-unitary operations that preserve the real structure of the state. Thus, the real part of the coherences obtained in the reconstructed density matrix, such as \(\ket{11}\bra{00}\), is at least fifty times larger than the imaginary part, which is consistent with the theoretical expectation for states devoid of complex phases. It is worth mentioning that the fidelities of the reconstructed matrices relative to the theoretical expectations are above 0.94, demonstrating the excellent performance of our experimental setup.  Given that all the optical elements of our setup implement near-unitary operations, there is no necessity to include decoherence channels in our theoretical description, reducing the complexity of our model.

One can see that an entangled state emerges when Alice performs projections onto the state $\ket{+}$. When $\theta=\pi/2$, the experiment gives a maximal entangled state with fidelity of 0.98.

%
%
\section{Avoiding retro-inference}
\label{app:B}

Commonly used in discussions of dual behavior in interferometric setups involving delayed choices~\cite{Auccaise2012,Peruzzo2012,Kaiser2012,Kim2012,Ma2016}, retro-inference consists of inferring the behavior of a quantum system $Q$ inside the interferometer, at past instants of time, based on measurement results obtained outside the interferometer (at later times). In what follows, we will use a Mach-Zehnder interferometer (MZI) to illustrate how our proposal of employing extra degrees of freedom (d.o.f.) $d_{1,2}$ can avoid the various conceptual problems associated with retro-inference.

\begin{figure*}[h]
\centering
\includegraphics[scale=0.8]{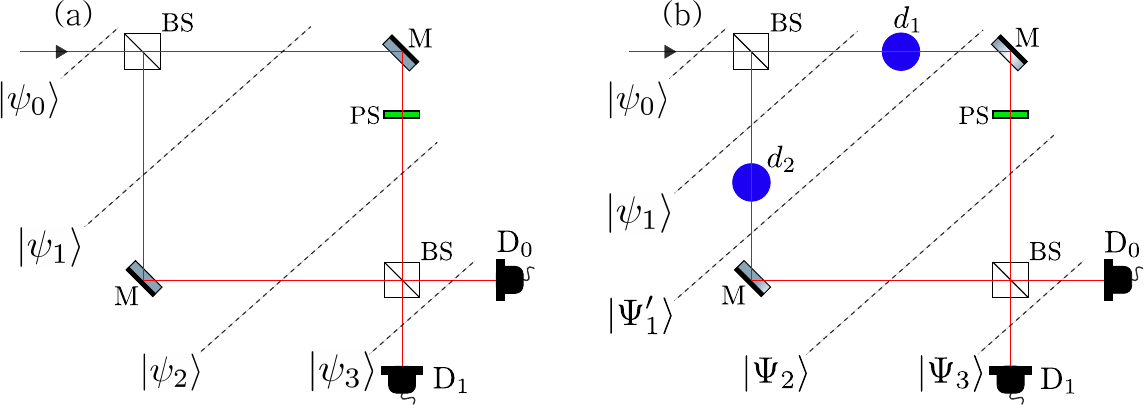}
\caption{(a) Usual structure of a Mach-Zehnder interferometer, with two beam-splitters (BS), two mirrors (M), a phase shifter (PS), and two detectors (D$_{0,1}$). (b) Two degrees of freedom (d.o.f.) $d_{1,2}$ are placed in the arms of the MZI so as to mark the path taken by $Q$. }
\label{fig:MZI}
\end{figure*}

In the configuration depicted in Fig.~\ref{fig:MZI}-(a), $Q$ enters the MZI in the state $\ket{\psi_0}=\ket{0}$ and, upon the action of BS, has its state transformed into 

\begin{equation}\label{psi1}
\ket{\psi_1}=\frac{1}{\sqrt{2}}\left(  \ket{0}+i \ket{1}\right),
\end{equation}
where $\ket{0}$ and $\ket{1}$ denote the horizontal and vertical spatial mode of $Q$, respectively. The mirrors M change the spatial mode and add a phase of $e^{i\pi/2}$, while the phase shifter PS introduces $e^{i\phi}$ in the respective path, hence $\ket{\psi_2}=(ie^{i\phi
}\ket{1}-\ket{0})/\sqrt{2}$.
After the second BS, one finds, up to some global phases,
\begin{align}\label{psi3}
\ket{\psi_3}  
=\cos\left(\tfrac{\phi}{2}\right)\ket{0} +\sin\left(\tfrac{\phi}{2}\right)\ket{1}.
\end{align}
The probabilities of clicks in the detectors D$_0$ and D$_1$ are given, respectively, by $p(0|\psi_3)\equiv |\braket{0|\psi_3}|^2=\cos^2(\phi/2)$ and $p(1|\psi_3)\equiv|\braket{1|\psi_3}|^2=\sin^2(\phi/2)$, so that the visibility $V_0$ of D$_0$ turns out to be
\begin{equation}
V_0(\psi_3) \coloneqq \frac{\max_\phi p(0|\psi_3) -\min_\phi p(0|\psi_3)  }{\max_\phi p(0|\psi_3) + \min_\phi p(0|\psi_3)}=1.
\end{equation}
An identical result is obtained for D$_1$. Phase-dependent probabilities and maximum visibility are typically considered diagnostics of wave-like behavior. In simpler terms, this means that $Q$ has chosen to travel both arms of the interferometer. Consider now the \textit{open} configuration of the MZI (not shown in Fig.~\ref{fig:MZI}), in which the second BS is removed. The probabilities of clicks in D$_{0,1}$ now read as $p(0|\psi_2)\equiv |\braket{0|\psi_2}|^2=\frac{1}{2}=|\braket{1|\psi_2}|^2\equiv p(1|\psi_2)$, and the D$_0$ visibility reduces to $V_0^\text{(open)}(\psi_2)=0$. These results are considered diagnostics of particle-like behavior, according to which $Q$ would have traveled through a single arm of the interferometer.

Now, this framework admits some fundamental criticisms. First, even when the MZI is open, the state after the first BS is given by the quantum superposition in Eq. \eqref{psi1}. Associating this superposition state with particle-like behavior in the open configuration and with wave-like behavior in the closed configuration implies subscribing to some hidden variables interpretation, according to which the quantum state alone is not sufficient to describe the physics of the system. Second, this attitude also raises all types of retrocausal and nonlocal issues that permeate delayed-choice experiments, since one can conceive that the behavior of $Q$ after the first BS is somehow influenced by the second BS, which could eventually be placed far away in spacetime. Third, visibility $V_0(\psi_3)$ is measured after $Q$ has physically interacted with the second BS and is then used to (retro)infer $Q$'s behavior before such interaction. Beam splitters introduce quantum coherences that cannot be neglected. Finally, all this rationale can lead one to envisage violations of Bohr's complementarity principle. For more detailed discussions, the reader is referred to Refs.~\cite{Angelo2015,Dieguez2022,Starke2024}.

The framework of quantum irrealism, on the other hand, is immune to these issues since it conditions the elements of reality of the system solely to the quantum state at each moment in time. So, taking $\sigma_z$ as the physical quantity that encodes the information about the path [since $\sigma_z\ket{\nu}=(-1)^\nu\ket{\nu}$, with $\nu\in\{0,1\}$], we readily conclude that $\mathfrak{I}_{\sigma_z}(\psi_1)=1$, meaning that $Q$ cannot have a spatial element of reality after the first BS, regardless of the configuration of the MZI. Any influence of further devices enters the discussion through interaction inserted in the Schrödinger equation, which then prescribes a local dynamics for $\mathfrak{I}_{\sigma_z}$, without appealing to retrodiction. Since the irreality measure requires state tomography, the question then arises as to how one can experimentally demonstrate dual behavior inside the interferometer. Conception of a way to realize this task is one of the key contributions of our work.

For the setup under scrutiny here, our proposal can be illustrated in terms of Fig.~\ref{fig:MZI}-(b).  The basic idea is to diagnose the wavelike behavior of $Q$ through the system's ability to induce entanglement between non-interacting d.o.f. $d_{1,2}$. If $Q$ has spatial elements of reality (particle-like behavior), then it cannot ``touch'' $d_{1,2}$ simultaneously and, therefore, cannot activate entanglement between these degrees of freedom. The initial state of the composed system, comprising $Q$ and the two other qubits $d_{1,2}$, is given by $\ket{\Psi_0}=\ket{\psi_0}\otimes \ket{11}_{d_1d_2}$. This state is unitarily transformed into $\ket{\Psi_1}=\ket{\psi_1}\otimes\ket{11}_{d_1d_2}$ after $Q$ interacts with the first BS, and into
\begin{equation}
\ket{\Psi_1^{\,\prime}} =\frac{1}{\sqrt{2}%
}\left(  \ket{0}\ket{01}_{d_1d_2}+i\ket{1}\ket{10}_{d_{1}d_{2}}\right),
\end{equation}
upon the interaction between $Q$ and $d_{1,2}$. After the mirrors and the PS, the state becomes $\ket{\Psi_2}=\frac{1}{\sqrt{2}}\left(
ie^{i\phi}\ket{1}\ket{01}_{d_1d_2}-\ket{0}\ket{10}_{d_1d_2}\right)$. Finally, after the second BS we obtain, up to global phases,
\begin{align}
\ket{\Psi_3} 
&  =\frac{1}{\sqrt{2}}\left(  \ket{0}\ket{\varphi_+}_{d_1d_2}  +i \ket{1}\ket{\varphi_-}_{d_1d_2}\right),
\end{align}
where $\ket{\varphi_\pm}_{d_1d_2}\equiv \left(\ket{10}_{d_1d_2} \pm e^{i\phi}\ket{01}\right)/\sqrt{2}$ are entangled states. Post-selection on either $\ket{0}$ or $\ket{1}$ leaves $d_{1,2}$ in entangled states, either $\ket{\varphi_+}_{d_1d_2}$ or $\ket{\varphi_-}_{d_1d_2}$, respectively. It is just an exercise to show that had $\ket{\psi_1}$ suddenly decohered to $\frac{1}{2}\left(\ket{0}\!\bra{0}+\ket{1}\!\bra{1} \right)$ before the interaction with $d_{1,2}$, we would rather have obtained the classically correlated state $\frac{1}{2}\left(\ket{10}\!\bra{10}_{d_1d_2}+\ket{01}\!\bra{01}_{d_1d_2}\right)$ after the post-selection. One might argue that this framework is also plagued by retro-inference, since dual behavior assessment can only be performed \textit{a posteriori}, after entanglement is detected at the output of the MZI. However, this is not true. The crux is that we have established a one-to-one correspondence between the presence (respectively absence) of entanglement in the final state of $d_1d_2$ with the degree of spatial indefiniteness $\mathfrak{I}_{\sigma_z}(\psi_1)=1$ (respectively $\mathfrak{I}_{\sigma_z}\left(\mathbbm{1}/2\right)=0$) of $Q$ after the first BS. Moreover, the entire rationale is tightly attached to the quantum state, whose underlying dynamics is solely governed by the Schrödinger equation in a causal way; no aspects of hidden variables theory, nonlocality, and retrocausality are needed.



\begin{thebibliography}{99}

\bibitem{EPR1935} A. Einstein, B. Podolsky, and N. Rosen,
Can Quantum-Mechanical Description of Physical Reality Be Considered Complete?, 
Phys. Rev. {\bf 47}, 777 (1935).

\bibitem{Bohr1935} N. Bohr,
Can Quantum-Mechanical Description of Physical Reality be Considered Complete?
Phys. Rev. {\bf 48}, 696 (1935).

\bibitem{Scully1982} M. O. Scully and K. Drühl,
Quantum Eraser: A Proposed Photon Correlation Experiment Concerning Observation and “Delayed Choice” in Quantum Mechanics,
Phys. Rev. A {\bf 25}, 2208 (1982).

\bibitem{Kwiat1992} P. G. Kwiat, A. M. Steinberg, and R. Y. Chiao,
Observation of a ``quantum Eraser’’: A Revival of Coherence in a Two-Photon Interference Experiment,
Phys. Rev. A {\bf 45}, 7729 (1992).

\bibitem{Herzog1995} T. J. Herzog, P. G. Kwiat, H. Weinfurter, and A. Zeilinger,
Complementarity and the Quantum Eraser,
Phys. Rev. Lett. {\bf 75}, 3034 (1995).

\bibitem{Garisto1999} R. Garisto and L. Hardy,
Entanglement of Projection and a New Class of Quantum Erasers,
Phys. Rev. A {\bf 60}, (1999).


\bibitem{Walborn2002} S. P. Walborn, M. O. Terra Cunha, S. Pádua, and C. H. Monken, 
Double-Slit Quantum Eraser,
Phys. Rev. A 
{\bf 65}, 033818 (2002).


\bibitem{Zubairy2004} M. S. Zubairy, G. S. Agarwal, and M. O. Scully,
Quantum Disentanglement Eraser: A Cavity QED Implementation,
Phys. Rev. A {\bf 70}, 012316 (2004).


\bibitem{Roos2004} C. F. Roos, M. Riebe, H. Häffner, W. Hänsel, J. Benhelm, G. P. T. Lancaster, C. Becher, F. Schmidt-Kaler, and R. Blatt,
Control and Measurement of Three-Qubit Entangled States,
Science {\bf 304}, 1478 (2004).


\bibitem{Starke2023_2} D. S. S. Chrysosthemos, M. L. W. Basso, and J. Maziero,
Quantum simulation of the generalized-entangled quantum eraser and the related complete complementarity relations,
Phys. Scr. \textbf{98}, 035107 (2023).

\bibitem{Jacques2007}
V. Jacques, E. Wu, F. Grosshans, F. Treussart, P. Grangier, A. Aspect, J.-F. Roch,
Experimental Realization of Wheeler's Delayed-Choice Gedanken Experiment, 
Science {\bf 315} (5814), 966 (2007).

\bibitem{Ma2016} X. Ma, J. Kofler, and A. Zeilinger,
Delayed-choice gedanken experiments and their realizations,
Rev. Mod. Phys. \textbf{88}, 015005 (2016).

\bibitem{Bell1964} J. S. Bell, 
On the Einstein Podolsky Rosen paradox,
Phys. Phys. Fiz. {\bf 1}, 195 (1964).

\bibitem{Bell2001} J. S. Bell,
in {\it John S. Bell on the Foundations of Quantum Mechanics}, edited by M. Bell, K. Gottfried, and M. Veltman (World Scientific, Singapore, 2001), pp. 50-60.


\bibitem{Hensen2015} B. HensenH. Bernien, A. E. Dr\'eau, A. Reiserer, N. Kalb, M. S. Blok, J. Ruitenberg, R. F. L. Vermeulen, R. N. Schouten, C. Abell\'an {\it et al.},
Loophole-Free Bell Inequality Violation Using Electron Spins Separated by 1.3 Kilometres,
Nature {\bf 526}, 682 (2015).

\bibitem{Giustina2015} M. Giustina, M. A. M. Versteegh, S. Wengerowsky, J. Handsteiner, A. Hochrainer, K. Phelan, {\it et al.},
Significant-Loophole-Free Test of Bell’s Theorem with Entangled Photons,
Phys. Rev. Lett. {\bf 115}, 250401 (2015).

\bibitem{Shalm2015} L. K. Shalm, E. Meyer-Scott, B. G. Christensen, P. Bierhorst, M. A. Wayne, M. J. Stevens, T. Gerrits, S. Glancy, D. R. Hamel, M. S. Allman {\it et al.},
Strong Loophole-Free Test of Local Realism, Phys. Rev. Lett. {\bf 115}, 250402 (2015).

\bibitem{Hensen2016} B. Hensen, N. Kalb, M. S. Blok, A. E. Dr\'eau, A. Reiserer, R. F. L. Vermeulen, R. N. Schouten, M. Markham, D. J. Twitchen, K. Goodenough {\it et al.},
Loophole-Free Bell Test Using Electron Spins in Diamond: Second Experiment and Additional Analysis,
Sci. Rep. {\bf 6}, 30289 (2016).

\bibitem{Rauch2018} D. Rauch, J. Handsteiner, A. Hochrainer, J. Gallicchio, A. S. Friedman, C. Leung, B. Liu, L. Bulla, S. Ecker, F. Steinlechner {\it et al.},
Cosmic Bell Test Using Random Measurement Settings from High-Redshift Quasars,
Phys. Rev. Lett. {\bf 121}, 080403 (2018).

\bibitem{Li2018} M. H. Li, C. Wu, Y. Zhang, W. Z. Liu, B. Bai, Y. Liu, W. Zhang, Q. Zhao, H. Li, Z. Wang {\it et al.},
Test of Local Realism into the Past without Detection and Locality Loopholes,
Phys. Rev. Lett. {\bf 121}, 080404 (2018).

\bibitem{Ma2013} X. Ma, J. Kofler, A. Qarry, N. Tetik, T. Scheidl, R. Ursin, S. Ramelow {\it et al.},
Quantum Erasure with Causally Disconnected Choice,
Proc. Natl. Acad. Sci. U.S.A. {\bf 110}, 1221 (2013).

\bibitem{Elitzur2001} A. C. Elitzur and S. Dolev, Nonlocal Effects of Partial Measurements and Quantum Erasure,
Phys. Rev. A \textbf{63}, 1 (2001).

\bibitem{Brodutch2016} A. Brodutch and E. Cohen, Nonlocal Measurements via Quantum Erasure, Phys. Rev. Lett. \textbf{116}, 070404 (2016).

\bibitem{Xu2019} X.-Y. Xu et al., Measurements of Nonlocal Variables and Demonstration of the Failure of the Product Rule for a Pre- and Postselected Pair of Photons, Phys. Rev. Lett. \textbf{122}, 100405 (2019).

\bibitem{Quintino2014} M. T. Quintino, T. V\'ertesi, and N. Brunner, Joint Measurability, Einstein-Podolsky-Rosen Steering, and Bell Nonlocality,
Phys. Rev. Lett. \textbf{113}, 160402 (2014).

\bibitem{Uola2014} R. Uola, T. Moroder, and O. G\"uhne, Joint Measurability of Generalized Measurements Implies Classicality, Phys. Rev. Lett. \textbf{113}, 160403 (2014).

\bibitem{Dieguez2022} P. R. Dieguez, J. R. Guimar\~aes, J. P. S. Peterson, R. M. Angelo, and R. M. Serra,
Experimental Assessment of Physical Realism in a Quantum-Controlled Device,
Commun. Phys. {\bf 5}, 82 (2022).

\bibitem{Bilobran2015} A. L. O. Bilobran and R. M. Angelo,
A Measure of Physical Reality,
Europhys. Lett. {\bf 112}, 40005 (2015).

\bibitem{Baumgratz2014} T. Baumgratz, M. Cramer, and M. B. Plenio,
Quantifying Coherence,
Phys. Rev. Lett. {\bf 113}, 140401 (2014).

\bibitem{Ollivier2001} H. Ollivier and W. H. Zurek, 
Quantum Discord: A Measure of the Quantumness of Correlations,
Phys. Rev. Lett. \textbf{88},  017901 (2001).

\bibitem{Rudnicki2018} \L{}. Rudnicki, Uncertainty-reality complementarity and entropic uncertainty relations, J. Phys. A: Math. Theor. \textbf{51}, 504001 (2018).

\bibitem{Dieguez2018} P. R. Dieguez and R. M. Angelo,
Information-Reality Complementarity: The Role of Measurements and Quantum Reference Frames,
Phys. Rev. A {\bf 97}, 022107 (2018).

\bibitem{Freire2019} I. S. Freire and R. M. Angelo,
Quantifying Continuous-Variable Realism,
Phys. Rev. A {\bf 100}, 022105 (2019).

\bibitem{Lustosa2020} F. R. Lustosa, P. R. Dieguez, and I. G. Da Paz, Irrealism from fringe visibility in matter-wave double-slit interference with initial contractive states,
Phys. Rev. A \textbf{102}, 052205 (2020).

\bibitem{Savi2021} M. F. Savi and R. M. Angelo,
Quantum Resource Covariance,
Phys. Rev. A {\bf 103}, 022220 (2021).

\bibitem{Gomes2022} V. S. Gomes, P. R. Dieguez, and H. M. Vasconcelos, Realism-based nonlocality: Invariance under local unitary operations and asymptotic decay for thermal correlated states, Physica A \textbf{601}, 127568 (2022).

\bibitem{Orthey2022} A. C. Orthey Jr. and R. M. Angelo,
Quantum Realism: Axiomatization and Quantification,
Phys. Rev. A {\bf 105}, 052218 (2022).

\bibitem{Basso2022} M. L. W. Basso and J. Maziero,
Reality Variation under Monitoring with Weak Measurements,
Quantum Inf. Process. \textbf{21}, 255 (2022).

\bibitem{Paiva2023} I. L. Paiva, P. R. Dieguez, R. M. Angelo, and E. Cohen,
Coherence and Realism in the Aharonov-Bohm Effect,
Phys. Rev. A {\bf 107}, 032213 (2023).

\bibitem{Engelbert2023} N. G. Engelbert and R. M. Angelo,
Considerations on the Relativity of Quantum Irrealism,
Entropy {\bf 25}, 603 (2023).

\bibitem{Orthey2025} A. C. Orthey Jr. and A. Streltsov,
Geometric Monotones of Violations of Quantum Realism, arXiv:2412.11633.


\bibitem{Orthey2025_2} A. C. Orthey Jr., P. R. Dieguez, O. Makuta, and R. Augusiak, High-dimensional monitoring and the emergence of realism via multiple observers, Phys. Rev. A \textbf{111}, 012220 (2025).



\bibitem{Mancino2018} L. Mancino, M. Sbroscia, E. Roccia, I. Gianani, V. Cimini, M. Paternostro, and M. Barbieri,
Information-Reality Complementarity in Photonic Weak Measurements,
Phys. Rev. A {\bf 97}, 062108 (2018).

%
%
\bibitem{Reid2017} M. D. Reid, Interpreting the macroscopic pointer by analysing the elements of reality of a Schr\"odinger cat, J. Phys. A: Math. Theor. \textbf{50}, 41LT01 (2017).


\bibitem{Moreira2019} S. V. Moreira and M. T. Cunha, Quantifying quantum invasiveness, Phys. Rev. A 99, 022124 (2019).


\bibitem{Zhu2021} H. Zhu, Zero uncertainty states in the presence of quantum memory, Npj Quantum Inf \textbf{7}, 47 (2021).


\bibitem{Thenabadu2022} M. Thenabadu and M. D. Reid, Macroscopic delayed-choice and retrocausality: quantum eraser, Leggett-Garg and dimension witness tests with cat states, Phys. Rev. A \textbf{105}, 062209 (2022).

\bibitem{Starke2024} D. S. Starke, M. L. W. Basso, and J. Maziero, An updated quantum complementarity principle, Proc. R. Soc. A \textbf{480}, 20240517 (2024).


\bibitem{Angelo2015} R. M. Angelo and A. D. Ribeiro,
Wave-particle duality: an information-based approach,
Found. Phys. \textbf{45}, 1407 (2015).

\bibitem{Starke2023_1} D. S. S. Chrysosthemos, M. L. W. Basso, and J. Maziero,
Quantum coherence versus interferometric visibility in a biased Mach–Zehnder interferometer,
Quantum Inf. Process. \textbf{22}, 68 (2023).

\bibitem{kwiat99} P. G. Kwiat et al., 
Ultrabright source of polarization-entangled photons,
Phys. Rev. A {\bf60}, R773 (1999).




\bibitem{James2001} D. F. V. James, P. G. Kwiat, W. J. Munro, A. G. White, 
Measurement of qubits,
Phys. Rev. A {\bf64}, 52312 (2001).





\bibitem{Auccaise2012} R. Auccaise, R. M. Serra, J. G. Filgueiras, R. S. Sarthour, I. S. Oliveira, L. C. Celeri, Experimental analysis of the quantum complementarity principle, Phys. Rev. A {\bf 85}, 032121 (2012).

\bibitem{Peruzzo2012} A. Peruzzo, P. Shadbolt, N. Brunner, S. Popescu, J. L. O’Brien, A quantum delayed-choice experiment, Science {\bf 338}, 634 (2012).

\bibitem{Kaiser2012} F. Kaiser, T. Coudreau, P. Milman, D. B. Ostrowsky, S. Tanzilli, Entanglement-enabled delayed-choice experiment, Science {\bf 338}, 637 (2012).

\bibitem{Kim2012} M. S. Kim, J. Lee, Y. J. Park, S. K. Choi, H. Nha, H. J. Kim, W. J. Munro, K. Nemoto, S. J. Kim, P. L. Knight, and J. Lee, Realization of Quantum Wheeler's Delayed-Choice Experiment, Phys. Rev. Lett. {\bf 108}, 230404 (2012).


\end{thebibliography}
\end{document}